\documentclass[10pt,twocolumn]{IEEEtran}

\usepackage{graphicx}           % Important!!
\usepackage{amsmath,amsfonts}
\usepackage{url}
\usepackage{times,epsfig}
\usepackage{psfrag}
\usepackage{subfigure}
\usepackage{stfloats}
\usepackage{amssymb}

\begin{document}

\title{Securing Cognitive Radio Networks against Primary User Emulation Attacks}

\author{Rong Yu, \emph{Member, IEEE}, Yan Zhang, \emph{Senior Member, IEEE}, Yi Liu, \emph{Member, IEEE}, \\Stein Gjessing, \emph{Senior Member, IEEE}, Mohsen Guizani, \emph{Fellow, IEEE}

\thanks{Rong Yu is with Guangdong University of Technology, China. Email: yurong@ieee.org}
\thanks{Yan Zhang and Stein Gjessing are with Simula Research Laboratory, Norway; Department of Informatics, University of Oslo, Norway. Email: yanzhang@ieee.org, steing@ifi.uio.no}
\thanks{Yi Liu is with Guangdong University of Technology, China, and Singapore University of Technology and Design, Singapore. Email: liuii5115@yahoo.com.cn}
\thanks{Mohsen Guizani is with Qatar University, Qatar. Email: mguizani@ieee.org} }

\maketitle \IEEEpeerreviewmaketitle

\begin{abstract}
Cognitive Radio (CR) is a promising technology for next-generation wireless
networks in order to efficiently utilize the limited spectrum resources and
satisfy the rapidly increasing demand for wireless applications and services.
Security is a very important but not well addressed issue in CR networks. In
this paper we focus on  security problems arising from Primary User Emulation
(PUE) attacks in CR networks. We present a comprehensive introduction to PUE
attacks, from the attacking rationale and its impact on CR networks, to
detection and defense approaches. In order to secure  CR networks against PUE
attacks, a two-level database-assisted detection approach is proposed to
detect such attacks. Energy detection and location verification are combined
for fast and reliable detection. An admission control based defense approach
is proposed to mitigate the performance degradation of a CR network under a
PUE attack. Illustrative results are presented to demonstrate the
effectiveness of the proposed detection and defense approaches.
\end{abstract}


\section{Introduction}

Cognitive Radio (CR) is  an enabling technology to effectively address the
spectrum scarcity and it will significantly enhance the spectrum utilization
of future wireless communications systems. In a CR network, the Secondary (or
unlicensed) User (SU) is allowed to opportunistically access the spectrum
``holes'' that are not occupied by the Primary (or licensed) User (PU).
Generally, the SUs constantly observe the spectrum bands by performing
spectrum sensing. Once a spectrum ``hole'' is discovered, an SU could
temporarily transmit on this part of the spectrum. Upon the presence of a PU
in this part of the spectrum, however, the SU has to switch to another
available spectrum band by performing spectrum handoff, avoiding interference
with the PU transmission.
The development of CR technology leads to the new communications paradigm
called Dynamic Spectrum Access (DSA), which relaxes the traditional fixed
spectrum assignment policy and allows a CR networks to temporally ``borrow''
a part of the spectrum from the primary network. As a consequence, the scarce
spectrum resources are shared, in a highly efficient and resilient fashion,
between the primary network and the CR network.

Among all the key technical problems of CR networks, security is a crucial
but not well addressed issue.
Due to the nature of dynamic spectrum access and the fact that the CR network
should not interact with the primary network, the SUs in the CR network
usually lack global information about the usage of the spectrum resource in
the network. This makes the CR network vulnerable to attacks by hostile
users.
In all the main functionalities of CR networks such as spectrum sensing,
spectrum mobility, spectrum sharing and spectrum management, the CR network
has been shown to be strategically vulnerable \cite{Zargar.JSAC}. The typical
attacks on CR networks may include Denial of Service (DoS) attacks, system
penetration, repudiation, spoofing, authorization violation, malware
infection, and data modification. These attacks cause potential threats to
the information confidentiality, integrity and availability of the CR
network. Effective defense approaches are urgently needed to  secure CR
networks and deal with these attacks. Nowadays, security threats and their
countermeasures have been studied as one of the most important topics in the
research area of CR technology  \cite{ClancyCrowncom2008}.

In this paper, we mainly focus on the security problem arising from Primary
User Emulation (PUE) attacks in CR networks.
PUE attacks are known as a new type of attacks unique to CR networks. In such
an attack, the hostile user takes the advantage of the inherent etiquette in
CR networks that the legitimate SU has to evacuate the spectrum band upon the
arrival of a  PU. An attacker emulates the PU's transmitting signal and
misleads the legitimate SU to give up the spectrum band. The presence of PUE
attacks may severely influence the performance of CR networks.
This paper aims at presenting a comprehensive introduction to PUE attacks,
from the attacking principle and its impact on CR networks, to the detection
and defense approaches. In order to secure  CR networks, we propose a
database-assisted detection approach and an admission control based defense
approach against PUE attacks.

\begin{figure*}[t]
\centerline{\includegraphics[width=0.78\textwidth]{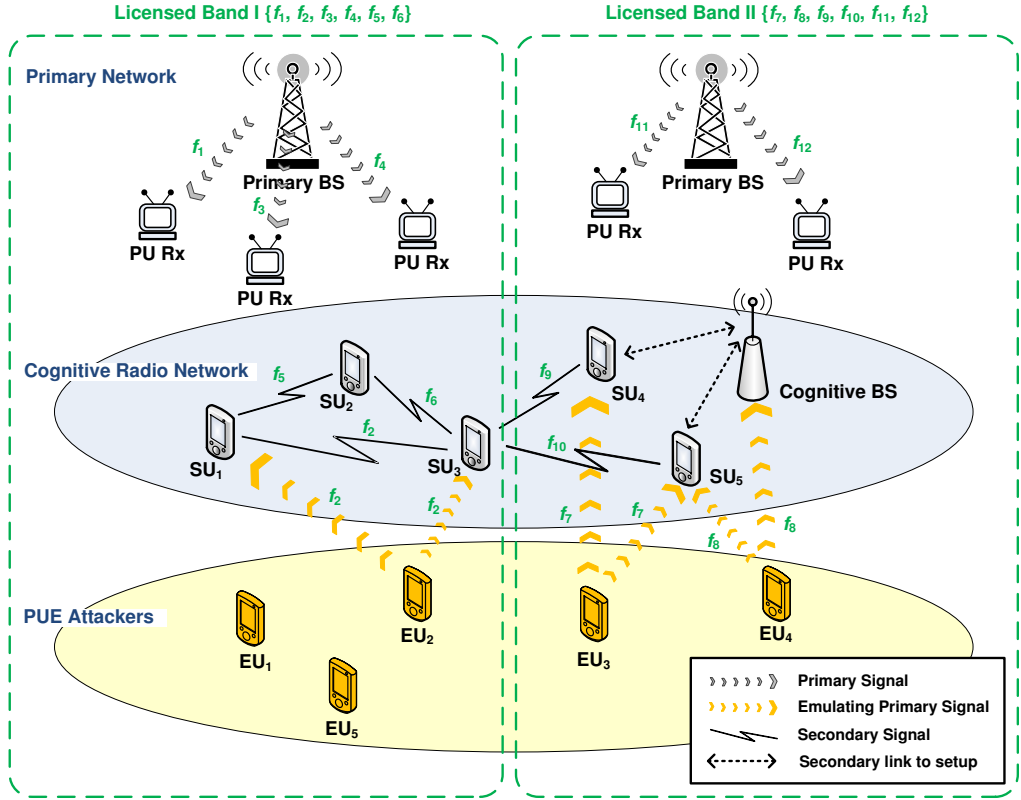}}
\centering\caption{Illustration of PUE attacks in the CR
network.}\label{fig:PUEA}
\end{figure*}

The remainder of the paper is organized as follows. Section II illustrates
the principles of PUE attacks, and introduces its classification and impacts
on  CR networks. A quantitative analysis of the performance degradation of a
CR network due to a PUE attack is also presented. Section III describes
existing detection measures for PUE attacks. A two-level database-assisted
detection approach is proposed. Energy detection and location verification
are combined for both fast and reliable detection. Section VI discusses the
defense approaches against PUE attacks, where a guard channel based admission
control is adopted to defend against PUE attacks. Finally, the conclusions of
the paper are presented in Section V.

\section{PUE Attack and Its Impact on CR Networks}

The term Primary User Emulation (PUE) attack was first introduced in
\cite{Chen06}. A PUE attack is a new type of attack unique to CR networks, in
which the attackers may modify their radio transmission frequency to mimic a
primary signal, thereby misguiding the legitimate SUs to erroneously identify
the attackers as a PU.

Fig.~\ref{fig:PUEA} shows a typical scenario of a PUE attack. There are two
spectrum bands, say, licensed band I and band II. Both of the spectrum bands
have six channels, indexed by frequencies $f_1, f_2,\cdots ,f_6$, and $f_7,
f_8,\cdots ,f_{12}$, respectively. Let's consider the first example in band
I, where the primary base station (BS) is transmitting in channels $f_1, f_3$
and $f_4$ to the PU receivers. Channels $f_2, f_5$ and $f_6$ are idle. By
observing this,  SU$_1$, SU$_2$ and SU$_3$ are allowed to use these three
idle channels for transmissions. However, the appearance of a PUE attacker,
say, EU$_2$, may block the SUs from using an idle channel. EU$_2$, may, for
example,  mimic the primary signal in channel $f_2$. Once the attack
succeeds, SU$_1$ and SU$_3$ are misled to evacuate channel $f_2$ and the link
between them is interrupted.
The second example is shown in band II. The primary network is occupying
channels $f_{11}$ and $f_{12}$, while  SU$_4$ and SU$_5$ are using channels
$f_{9}$ and $f_{10}$, respectively. PUE attackers EU$_3$ and EU$_4$ are
emulating the primary signals in channels $f_7$ and $f_8$, respectively. In
this situation, suppose that  SU$_4$ and SU$_5$ need to find channels to
connect with the cognitive base station (BS). If attackers EU$_3$ and EU$_4$
can not be correctly identified,  SU$_4$ and SU$_5$ will find no vacant
channels and hence may not be able to communicate with the cognitive BS.

The above two examples describe two different attacking cases.
The first example illustrates the case that the PUE attacker attacks the
in-service SUs and seizes one of their channels, causing interruption of some
of the  SU services.
The second example illustrates the case that the PUE attackers occupy the
idle channels and waste the spectrum opportunities of the SUs.

\subsection{Classification of Attackers}

Since the security problem caused by PUE attacks was identified, different
types of PUE attacks have been studied. We now introduce different types of
PUE attackers associated with their classification criteria.

\begin{itemize}
  \item \emph{Selfish \& Malicious Attackers:}
  A selfish attacker aims at stealing bandwidth from legitimate SUs
  for its own transmissions. The attacker will monitor the spectrum.
Once an unoccupied spectrum band is discovered, it will compete with the
legitimate SUs by emulating the primary signal, e.g.,  SU$_3$ and SU$_4$ in
Fig.~\ref{fig:PUEA}. A selfish attacker is a rational attacker in the sense
that, if it is detected by the legitimate SUs and the SUs reclaim the
spectrum opportunity by switching back to the band, it has to leave the band.
The purpose of a malicious attacker, however, is to disturb the dynamic
spectrum access of legitimate SUs but not to exploit the spectrum for its own
transmissions. Being different from a selfish attacker, the malicious
attacker may emulate a primary signal in both an unoccupied spectrum band and
a band currently used by legitimate SUs, e.g.,  SU$_2$ in
Fig.~\ref{fig:PUEA}. When an attacker attacks a band being used by a
legitimate SU, there exists the possibility that the SU fails to discover the
signal, and hence, an interference occurs between the attacker and the
legitimate SU.
  \item \emph{Power-Fixed \& Power-Adaptive Attackers:} The ability to
  emulate the power levels of a primary signal is crucial for PUE
  attackers, because most of the SUs employ an energy detection technique
  in spectrum sensing. A power-fixed attacker uses an invariable predefined
  power level regardless of the  actual transmitting power of the PUs and
  the surrounding radio environment.
      Compared to the power fixed attacker, the power-adaptive attacker
       is smarter in the sense that, it could adjust its transmitting
       power according to the estimated transmitting power of the primary signal and the channel parameters \cite{ZChen09}. Specifically, the attacker employs an estimation technique and a learning method against the detection by the legitimate SUs. It is demonstrated that such an advanced attack can defeat a naive defense approach that focuses only on the received signal power.
  \item \emph{Static \& Mobile Attackers:} The location of a signal source is also
  a key characteristic to verify the identity of an attacker. A static attacker has a fixed location that would not change in all round of attacks.
      By using positioning techniques such as Time of Arrival (ToA) or dedicated positioning sensors \cite{Chen08}, the location of a  static attacker could be revealed. A static attacker will be easily recognized due to the difference between its location and that of the PUs. A mobile attacker will constantly change its location so that it is difficult to trace and discover. A viable detection approach that exploits the correlations between RF signals and acoustic information is proposed in \cite{SChen11} to verify the
      existence of a mobile PUE attacker.
\end{itemize}

\subsection{Essential Conditions for Successful PUE Attacks}

In a CR network, the successful realization of a PUE attack relies on several
essential conditions. To better understand PUE attacks and facilitate the
design of the countermeasures, we summarize these essential conditions as
follows.

\begin{itemize}
  \item \emph{No PU-SU interaction:} There is no interaction between the primary and the secondary networks. This is a necessary condition for a successful PUE attack. Otherwise, if the legitimate SUs are allowed to exchange information with the PUs, a PU verification procedure could be designed to easily detect a PUE attack. In most cases, this condition holds. It is regulated in the IEEE 802.22 standard and also a general assumption in most
  existing research work of CR networks.
  \item \emph{PU and SU signals have different characteristics:} The primary
  and secondary networks use wireless signals with different
  characteristics, i.e., using different modulation modes and different signal statistical features. An SU receiver is inherently designed only for the secondary signal but unable to demodulate and decode the primary signal. The PUE attackers take advantage of this fundamental condition to emulate the primary signal that is unrecognisable for the legitimate SUs.
  \item \emph{Primary signal learning and channel measurement:} To emulate the primary signal, the attacker has to track and learn the characteristics of the primary signal. For an advanced attack, the attacker may also estimate the power level as well as the channel conditions to generate more tricky transmitting signals.
  \item \emph{Avoiding interference with the primary network:} Although this is usually a primary concern for the SUs, it is also a important condition that the PUE attackers have to comply with. The attackers, especially the selfish ones, should carefully monitor the behaviors of PUs as not to cause extra interference with the primary network.
\end{itemize}

\subsection{Impact of PUE attacks on CR Networks}

The presence of PUE attacks causes a number of troublesproblems for CR
networks. The list of potential consequences of PUE attacks is:

\begin{itemize}
  \item \emph{Bandwidth waste:} The ultimate objective of deploying CR networks is to address the spectrum under-utilization that is caused by the current fixed spectrum usage policy.
By dynamically accessing the spectrum ``holes'', the SUs are able to retrieve
these otherwise wasted spectrum resources. However, PUE attackers may steal
the spectrum ``holes'' from the SUs, leading to spectrum bandwidth waste
again.
  \item \emph{QoS degradation:} The appearance of a PUE attack may severely degrade the Quality-of-Service (QoS) of the CR network by destroying the continuity of secondary services. For instance, a malicious attacker could disturb the ongoing services and force the SUs to constantly change their operating spectrum bands. Frequent spectrum handoff will induce unsatisfying delay \cite{Jin12} and jitter for the secondary services.
  \item \emph{Connection unreliability:} If a realtime secondary service is attacked by a PUE attacker
and finds no available channel when performing spectrum handoff, the service
has to be dropped. This realtime service is then terminated due to the PUE
attack. In principle, the secondary services in CR networks inherently have
no guarantee that they will have stable radio resource because of the nature
of dynamic spectrum access. The existence of PUE attacks significantly
increases the connection unreliability of CR networks.
  \item \emph{Denial of Service:} Consider PUE attacks with high attacking frequency; then the attackers may occupy many of the spectrum opportunities. The SUs will have insufficient bandwidth for their transmissions, and hence, some of the SU services will be interrupted. In the worst case, the CR network may even find no channels to set up a common control channel for delivering the control messages. As a consequence, the CR network will be suspended and unable to serve any SU. This is called Denial of Service (DoS) in CR networks.
  \item \emph{Interference with the primary network:} Although a PUE attacker is motivated to steal the bandwidth from the SUs, there exists the chance that the attacker generates additional interference with the primary network. This happens when the attacker fails to detect the occurrence of a PU. On the other hand, when the SUs are tackling a PUE attack, it is also possible to incorrectly identify the true PU as the attacker and interfere with the primary network.
  In any case, causing interference with the primary network is strictly forbidden in CR networks.
\end{itemize}

\begin{figure}[t]
\centerline{\includegraphics[width=0.48\textwidth]{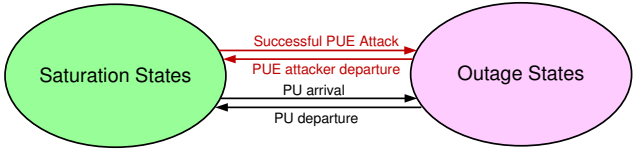}}
\centering\caption{Saturation state and outage state
transition.}\label{fig:emam}
\end{figure}

\subsection{Performance Degradation due to PUE Attacks}

We adopt the term \emph{saturation} to characterize the state of a CR network
in which all the channels are occupied by PUs, SUs and PUE attackers, i.e.,
there are no idle channels, and the term \emph{outage} to characterize the
state of a CR network in which there is no spectrum band available for the
Common Control Channel (CCC). In a practical CR network, it is necessary to
build up a CCC for exchanging control messages. The CCC might be established
by using a dedicated radio transceiver and setting up an out-of-band fixed
channel. However, this is very difficult in a real CR network due to the
additional cost of hardware and the assignment of a dedicated spectrum band.
It is more likely that the CCC should be constructed by means of dynamic
spectrum access. This implies that the CR network need to maintain a stable
channel as its CCC. Under PUE attacks, the CCC may also be attacked and
disconnected. The system will be suspended in this case. Two new performance
metrics are defined as follows.

\begin{itemize}
  \item \emph{Outage probability:} The outage probability is defined as the probability that a CR network stays in the outage state in
which there is no available spectrum band for constructing a CCC.
  \item \emph{System recovery time:} The system recovery time is defined as the average time duration that a CR network (in the outage state) takes to acquire an available spectrum band as a CCC for delivering control messages.
\end{itemize}

Fig.~\ref{fig:emam} shows the saturation state and the outage state
transition. When the current CCC is not available any more, due to the
arrival of a PU or a PUE attack, it has to switch to a new channel. Since the
CCC has priority over the other common SUs, it could switch to anyone of the
available channels, even one already occupied by the SUs. As a consequence,
the CCC is disconnected only in the case that all of the channels are
occupied by PUs or PUE attackers. In the saturation states with only one SU
channel being used as the CCC, if a PU arrives and occupies the CCC, or if a
PUE attacker successfully attacks the CCC, the CR network transits to the
outage state.

\begin{figure}[t]
%\centerline{\includegraphics[width=0.48\textwidth]{pics/recovery_time.eps}}
\centering
\begin{minipage}[b]{0.5\textwidth}
\includegraphics[width=1\textwidth]{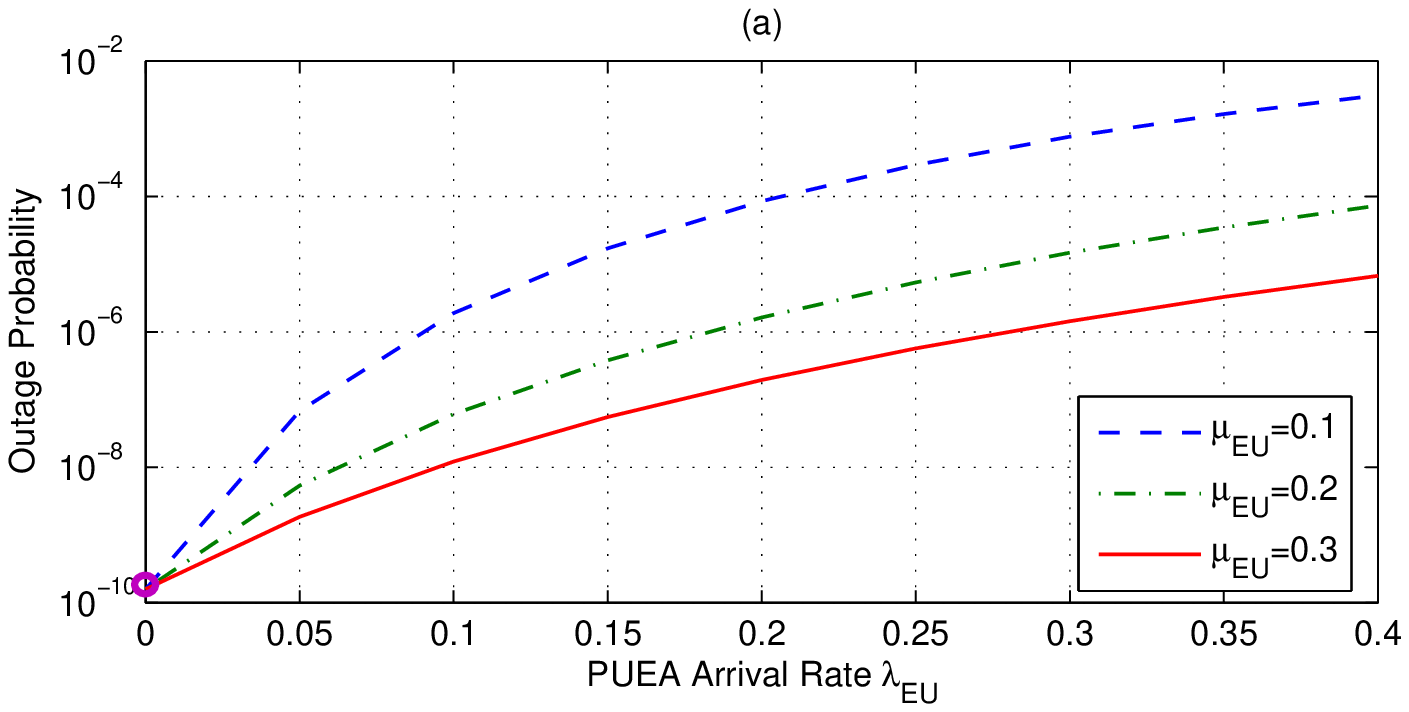}\\
\includegraphics[width=1\textwidth]{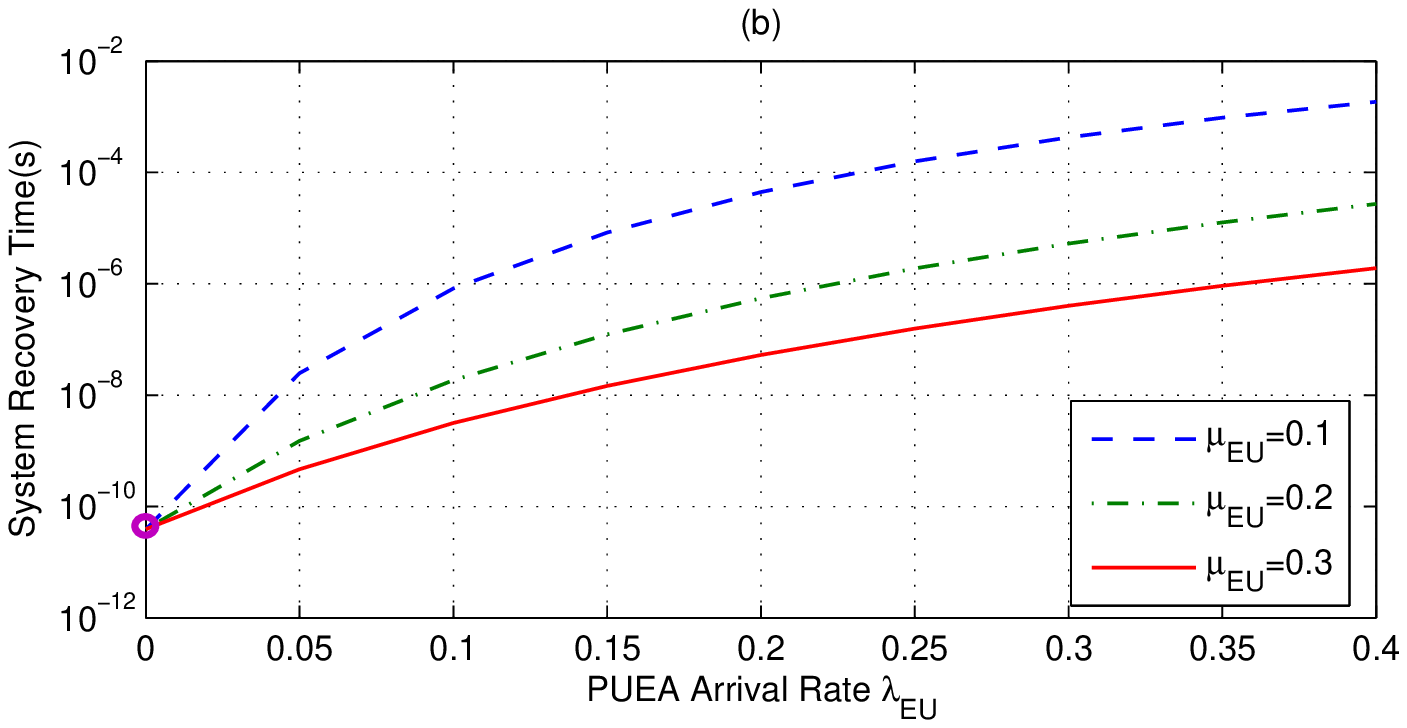}
\end{minipage}
\caption{Outage probability and system recovery time.}
\label{fig:recoverty_time}
\end{figure}

%\subsubsection{Outage probability and system recovery time}
%We consider a CR network with $N$ channels. Let ($i,j,k$) denote the system
%state where $i,j,k$ are the number of in-service PUs, SUs and PUE attackers,
%respectively. Let $p_{i,j,k}$ denote the steady state probability of the
%state $(i,j,k)$, and $\Omega$ the set of all states. The outage probability
%is the sum of the steady state probabilities of the outage states, i.e.,
%$P_{\mathrm{out}}=\sum_{j=0}p_{i,j,k}$. From the outage state, the system
%transits to the saturation state due to the departure of a  PU or a  PUE
%attacker, and transits to another outage state due to a PU arrival. Let
%$\lambda_{\mathrm{PU}}$, $\mu_{\mathrm{PU}}$ and $\mu_{\mathrm{EU}}$ denote
%the PU arrival and departure rates, and the PUE attacker departure rate,
%respectively. The time duration the system stays in the outage state is given
%by
%\begin{equation}
%T_{\mathrm{rec}}\!\!=\!\!\!\!\sum_{i<N,j=0}\!\frac{1}{i\mu_{\mathrm{PU}}+k\mu_{\mathrm{EU}}+\lambda_{\mathrm{PU}}}p_{i,j,k}+\frac{1}{N\mu_{\mathrm{PU}}}p_{N,0,0}.
%\end{equation}
%

%\subsubsection{Illustrative Result}

Fig.~\ref{fig:recoverty_time} shows the outage probability and the system
recovery time in terms of the PUE attacking strength, i.e., the attack
arrival rate. In the figure, $\lambda_{\mathrm{EU}}$ and $\mu_{\mathrm{EU}}$
denote the PUE attacker arrival rates and the PUE attacker departure rate,
respectively. It is observed that, both the outage probability and the system
recovery time increase dramatically with the increase of the attacking
strength. Without PUE attacks, the outage probability is near zero and the
recovery time is very short. In the case of a PUE attack, say,
$\lambda_{\mathrm{EU}}=0.4$ and $\mu_{\mathrm{EU}}=0.1$, the outage
probability is over $0.3\%$ and the recovery time is nearly 2ms. Hence, the
outage probability increases dramatically, and the recovery time extend
substantially, compared to the case when there are no PUE attacks. These
observations indicate that, the existence of PUE attacks may seriously
degrade the performance of a CR network. Detection and defense approaches
against PUE attacks are becoming very critical to secure CR networks.

\section{Detection Approaches for PUE Attacks}

\subsection{Existing Detection Approaches}

In the literature, some detection approaches against  PUE attacks have been
presented. The existing detection approaches can be classified into energy
detection, Received Signal Strength (RSS) based detection, feature detection,
location verification and cooperative detection.

\subsubsection{Energy Detection}

Energy detection is a simple but widely used approach for spectrum sensing in
CR networks. It is also one of the basic approaches for the detection of PUE
attacks. By measuring the power level of the received signal at the SU
receiver and comparing it with that from the true PUs, the CR network could
judge whether the signal comes from an attacker or not. However, a pure
energy detector is not robust enough to tackle an advanced PUE attack.

\subsubsection{RSS-based Detection}

Received Signal Strength (RSS) based detection approach is discussed in
\cite{Jin09}, where the authors analyze the PUE attack in the CR network
without using any location information. Thus, this detection approach does
not need dedicated sensor networks. The PUE attackers are assumed to be
distributed randomly around the SUs. The authors present an analysis using
Fenton¡¯s approximation and Wald¡¯s sequential probability ratio test (WSPRT)
to detect PUE attacks.

\subsubsection{Feature Detection}

The approach proposed in \cite{Pu11} uses energy detection to identify the
existing users in the frequency band. The approach then employs a
cyclostationary calculation to represent the features of the user signals,
which are then fed into an artificial neural network for classification. As
opposed to current techniques for detecting PUE attacks in CR networks, this
approach does not require additional hardware or time synchronization
algorithms in the wireless network.

\subsubsection{Location Verification}

Two location verification schemes are proposed in \cite{Chen06}. They are
called Distance Ratio Test (DRT) and Distance Difference Test (DDT),
respectively. In both schemes, dedicated cognitive nodes (SUs or a cognitive
BS) with enhanced functionality are involved for location verification. DRT
uses a Received Signal Strength (RSS) based method, where two dedicated
cognitive nodes measure the RSS of the signal source and calculate the ratio
of these two RSS to check whether it coincides with their distances to the
true PU (e.g., a TV broadcast tower). Using DDT, the arrival time of the
transmitted signal from the source is measured by the two dedicated cognitive
nodes. The product of the time difference and the light speed is then
compared to the distance difference from the true PU to the two dedicated
nodes in order to identify the source.

\begin{figure*}[t]
\centerline{\includegraphics[width=10cm]{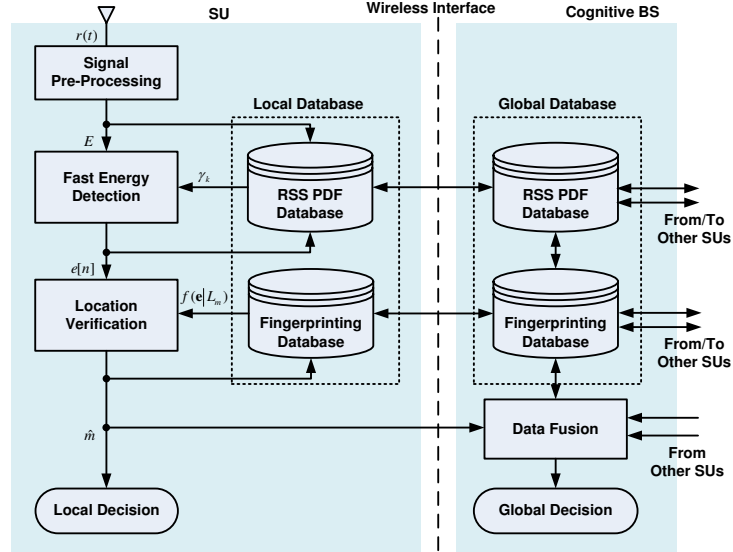}}
\centering\caption{Proposed database-assisted detection
approach.}\label{fig:DADA}
\end{figure*}

\subsection{A Database-Assisted Detection Approach}

Fig.~\ref{fig:DADA} shows our proposed database-assisted PUE attack detection approach, which has three key components: the multi-threshold fast energy detection, the fingerprint-based location verification and the two-level database.
In the approach, a local database is integrated in each SU, while a global database is built up in the cognitive BS. The local database is used to store historical spectrum sensing data and the local detection decisions of each SU.
The global database is used to collect and record all the SUs' spectrum
sensing data and the local detection decisions, as well as the global
detection decisions. If the proposed approach is applied in wireless regional
area networks, according to the IEEE 802.22 standard \cite{802.22}, the
global database in a cognitive BS can provide an interface to the incumbent
database for information query, e.g., the geo-location of a primary BS and
the list of available channels.
The main operations of the proposed detection approach are explained as follows.

\subsubsection{Basic Operations}

We consider a system model in which there are one primary BS (e.g., the TV
broadcasting tower) and multiple PUE attackers. In our model the attackers
are static or quasi-static, say, moving very  slowly. In a given moment and
in a specific spectrum band (channel), only one of the attackers, at most, will emulate a  primary signal.

In the proposed approach, there are four main units in the SU: a signal
pre-processing unit, a fast energy detector, a location verifier and a local
database. The local database consists of two components: An RSS Probability
Density Function (PDF) database and a fingerprint database.
The signal pre-processing unit gets the received signal $r(t)$ from the radio
frequency (RF) unit as input. Let $x(t)$, $h(t)$ and $\omega(t)$ denote the
transmitted signal, channel impulse response and the receiver thermal noise,
respectively. Let $s(t)$ and $s'(t)$ denote the real PU signal and the PUE
attack signal, respectively. Then, the transmitted signal $x(t)=s(t)$ for the
real PU signal, $x(t)=s'(t)$ for the PUE attack signal, and $x(t)=0$ when no
signal is transmitted. The input signal is given by $r(t)=x(t)\ast
h(t)+\omega(t)$. Let $\{t_n\}$ denote the sequence of sampling times and
$N_s$ the number of samples in one sensing period. After sampling, squaring
and aggregation, the signal pre-processing unit generates the sampled energy
vector $\mathbf{e}=e[n]$ ($n=1,2,\cdots,N_s)$ and the aggregated energy $E$.
Then, we have the sampled energy $e[n]=r^2(t_n)$ and the aggregated energy
$E=\sum_1^{N_s} e[n]$. After that, the aggregated energy $E$ is sent to the
fast energy detector for comparison to the preset thresholds $\gamma_k$'s. If
the comparison result indicates that there is no signal or it is a PUE attack
signal, the detection procedure is terminated and the corresponding decision
is made. Otherwise, the energy vector $e[n]$, containing more detailed energy
information, is sent to the location verification unit. The location of the
source of the signal is estimated using Bayesian hypothesis testing. The
estimated location $\hat{m}$ of the signal source is then transmitted to the
cognitive BS for data fusion. The operations of fast energy detector and
location verifier are elaborated below.

\subsubsection{Multiple Thresholds based Fast Energy Detection}

The goal of a fast energy detector is to quickly react to possible PUE
attacks.
The basic idea of a fast energy detector stems from conventional energy
detection.
In a conventional energy detector, there is only one energy threshold, to
distinguish the cases of presence or absence of a primary signal.
This single-threshold detector is not efficient for detecting a  PUE attack
signal.
To distinguish a PUE attacker from a real PU, a fast energy detector sets up
three energy thresholds, denoted by $\gamma_0, \gamma_1$ and $\gamma_2$.
Here, $\gamma_0<\gamma_1<\gamma_2$, and $\gamma_0$ is according to the
original threshold in a conventional energy detector. If the input
$E<\gamma_0$, it is decided that there is no PU or  PUE attacker present.
The two new thresholds $\gamma_1$ and $\gamma_2$ are used to distinguish the
signals of PU and PUE attacker. If the input $\gamma_0<E<\gamma_1$ or
$E>\gamma_2$, it is decided that a PUE attack is detected. Otherwise, the
received signal is initially diagnosed to be a PU signal. The location
verifier will be launched for further examination.
It is emphasized that, using two energy thresholds to distinguish a PU from a PUE attacker is justified by the following fact.
A PUE attacker tries to emulate the transmitting power of a real PU.
However, it is very difficult for the attacker to fabricate a signal so that
all of the SUs receive the signal with the power level similar to that of the
real PU.
By randomly assigning a few SUs to measure the received signal power, and
letting these SUs know the signal power of the real PUs, a PUE attack could
be discovered with a high probability.

Generally, the received energy $E$ has the form of a Chi-Square distribution.
Since the number of samples is large in most cases, we can use the Central
Limit Theorem (CLT) to approximate the Chi-Square distribution by a Gaussian
distribution.
Let $H_0, H_1$ and $H'_1$ denote the hypothesis of receiving no signal, a
real PU signal and a PUE attack signal, respectively.
Let $P_{d}(\gamma_1,\gamma_2)$ and $P_{f}(\gamma_1,\gamma_2)$ denote the PUE
attack detection and false alarm probabilities, respectively.
We have $P_{d}(\gamma_1,\gamma_2)=\Pr \{\gamma_0<E<\gamma_1|H'_1\}+\Pr
\{E>\gamma_2|H'_1\}$, and $P_{f}(\gamma_1,\gamma_2)=\Pr
\{\gamma_0<E<\gamma_1|H_1\}+\Pr \{E>\gamma_2|H_1\}$.
%
%The optimal thresholds $\gamma_1$ and $\gamma_2$ can be derived by solving
%the following constrained optimization problem:
%%
%\begin{eqnarray}\label{th_optimize}
%\max_{\gamma_1,\gamma_2}  & \quad &
%P_{d}(\gamma_1,\gamma_2)\quad \quad \\
%\mathrm{s.t.} & \quad & P_{f}(\gamma_1,\gamma_2)\leq \eta \nonumber
%\end{eqnarray}
%%
%where $\eta$ is the constraint of the PUE attack false alarm probability. In
%(\ref{th_optimize}), the PUE attack false alarm probability is constrained
%because the PUE attack false alarm should be stringently limited. When a PUE
%attack false alarm is triggered, the real PU is erroneously recognized as the
%attacker and the SUs will  not give up the spectrum band. Severe interference
%with the primary network may be induced in this case.
%%
%We also emphasize that, the RSS PDF from the local database will
%significantly facilitate the computation of $P_d$ and $P_f$ in
%(\ref{th_optimize}). The initial detection decision as well as the aggregated
%energy are fed back to the local database for updating the RSS PDF.
%

\subsubsection{Data Fusion driven Location Verification}

The proposed location verification does not need any dedicated positioning
sensors \cite{Chen08}. In particular, suppose that the global database has
recorded the location fingerprints of $M$ PUE attackers as well as that of
the real PU. The location verification will specifically identify the source
of the received signal from the real PU and the PUE attackers.
The location verification consists of three main steps. In step one, the SUs
observe the input energy vector $\mathbf{e}$ and estimate the location of
the source by finding the best matching entry in their local databases.
In step two, the SUs send the estimated location to the cognitive BS for data
fusion. The cognitive BS makes a final decision and identifies the signal
source.
In step three, the cognitive BS updates the global database according to the
gathered fingerprinting information from the multiple SUs. An update
information is also sent to the SUs' local databases.

The location estimation using Bayesian hypothesis testing is described as
follows.
Let $L_m$ ($m=0,1,2,\cdots,M$) denote the location of the signal source,
where $L_0$ corresponds to the real PU and $L_{\{1,2,\cdots,M\}}$ correspond
to the attackers, respectively.
The input energy vector $\mathbf{e}$ follows a parameterized probability
density function with the parameter  stored in the database.
Specifically, the probability density function of $\mathbf{e}$ under the
hypothesis that the source is located in $L_m$ is denoted
$f(\mathbf{e}|L_m)$. The estimation of the location of the  source of the
signal is given by
\begin{equation}\label{bayes}
\hat{m}= \mathop{\arg\max}_{m=0,1,2,\cdot,M}\pi_m f(\mathbf{e}|L_m)
\end{equation}
where $\pi_m$ is the a priori probability of the hypothesis that the source
is located in $L_m$.

The estimated location $\hat{m}$ is sent as the local decision to the
cognitive BS for data fusion. The data fusion rules that lead to various
global decisions are explained below.
\begin{itemize}
  \item \emph{True PU:} If all local decisions are identical and $L_0$, i.e. $\hat{m}=0$, the cognitive BS will decide that the signal source is the true PU.
  \item \emph{PUE attack in a known location:} If all local decisions are identical and
$\hat{m} \in L_{\{1,2,\cdots,M\}}$, the cognitive BS will decide that the
source is the PUE attacker in location $L_m$.
  \item \emph{PUE attack in a new location:} If the local decisions are different, the cognitive BS will decide that the source is a PUE attacker in a new location. A new fingerprint entry is added to the global database.
\end{itemize}

The final decision will be sent by the cognitive BS to the SUs. Both the
local and global databases will be updated when a PUE attack is detected,
either by the fast energy detector or by the location verifier. In
particular, in the fast energy detector, the energy thresholds will be
re-computed. In the location verifier, the probability density functions of
the energy vector will be updated. In addition, if a new location of a PUE
attacker is detected, a new profile will be created to track this new
attacker.
The communication overheads to update the two-level database is proportional to the frequency of PUE attacks.
The computational complexity in detecting PUE attacks is determined by the number of samples in each spectrum sensing and
the number of possible locations of the PUE attackers. The overall
computational complexity is $O(MN_s)$, which is sufficiently low for
practical deployment.

\begin{figure}[t]
\centerline{\includegraphics[width=0.49\textwidth]{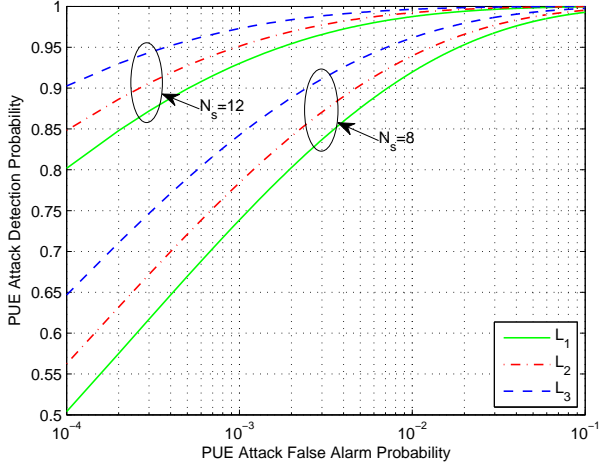}}
\centering\caption{PUE Attack detection probability in terms of false alarm
probability.}\label{fig:PUEA_detection}
\end{figure}

\subsubsection{Illustrative Result}

We consider a scenario where there are three PUE attackers located in
positions $L_1$, $L_2$ and $L_3$, respectively. The SUs are distributed in a
circular field with radius $1$ km. The primary BS is located in the center,
while $L_1$, $L_2$ and $L_3$ are respectively $100$ m, $200$ m, and $300$ m
away from the center. Fig.~\ref{fig:PUEA_detection} demonstrates the
effectiveness of the proposed detection approach. The PUE attack detection
probability is shown in terms of the false alarm probability. In this
example, we have shown two cases when the sampling parameter varies. The
comparison indicates that more samples lead to higher detection probability.
We can observe that, the farther the PUE attacker is located from the primary
BS, the easier it is to detect it. For example, when $N_s=12$ and
$P_f=0.1\%$, the PUE attack detection probabilities are $0.93$, $0.95$ and
$0.97$ when the PUE attacks are performed from locations $L_1$, $L_2$ and
$L_3$, respectively. The results indicate that the proposed approach works
effectively and is able to successfully detect the attacks.

\section{Defense Approaches against PUE Attacks}

The defense against PUE attacks is an important but seldom explored topic in
CR networks.
There are practical requirements for efficient PUE attack defense approaches.
We illustrate this by two examples below.
First, although a variety of PUE attack detection approaches have been
proposed, none of the existing  approaches is able to promise accurate
detection of all attacks. There still is a  chance that some attacks are not
detected. This necessitates system level mechanisms to maintain the overall
performance of a CR network under undetected PUE attacks.
Second, when there are malicious attackers in the network, their purpose is
to interrupt the communications of the cognitive users. Even if they have
been discovered,  malicious attackers may stil transmit in order to interfere
with the transmissions of the SUs. In this case, the signal processing units
in the RF front-ends of the SU receivers should be applied to get rid of the
interference signals, in order to  try to recover the secondary signal.

\subsection{Defense Approaches at Various Protocol Layers}

To defend against PUE attacks, effective counter measures could be taken at
different layers of the communications protocol stack.
\begin{itemize}
  \item \emph{Physical-layer approach:} Physical-layer techniques such as
  source separation, signal design, spread spectrum and directional antennas
   could be employed to deal with the intended interference from malicious PUE attackers.
      The key in the  design of an efficient physical-layer countermeasure
      is to exploit the a priori knowledge  about the characteristics of
      the primary signal and its dissimilarity with the interference signal.
  \item \emph{MAC-layer approach:} Undetected PUE attacks will steal bandwidth from the CR network. To let the SUs maintain moderate QoS performance, Radio Resource Management (RRM) strategies such as admission control, spectrum handoff and spectrum scheduling should be studied.
  \item \emph{Network-layer approach:} In cognitive ad hoc networks, once the
  location of the PUE attackers are estimated, a position-based cognitive routing
  strategy could be employed to deal with the PUE attacks. Those SUs that are
  located within the attacking range of the PUE attackers should be considered
  to be temporary unavailable. End-to-end routing paths should be established
  without crossing the unavailable SU nodes.
  \item \emph{Cross-layer approach:}
A cross-layer design framework may be set up to defend against PUE attacks.
In the framework, the behavior of the detected PUE attacks is observed at the
physical layer and reported to the upper layers, such as the RRM mechanism at
the MAC layer or the routing mechanism at the network layer. We emphasize
that, even the undetected PUE attacks could be estimated in the physical
layer by considering the theoretically derived detection probability. The
control parameters of the upper layer are jointly optimized considering the
existence of PUE attacks.
\end{itemize}

\begin{figure}
  \centering
  \includegraphics[width=0.5\textwidth]{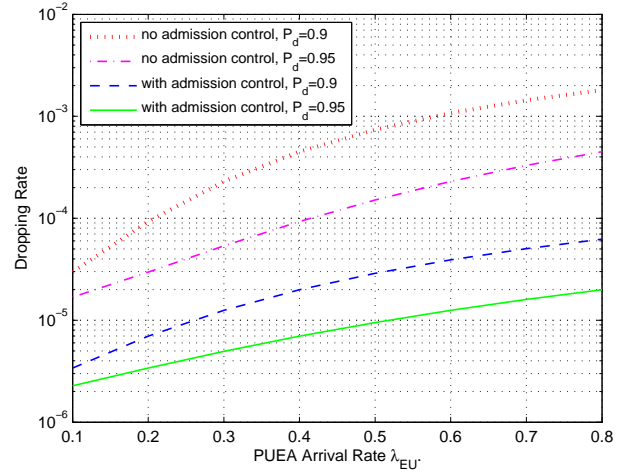}
  \caption{Performance comparison in dropping rate of CR networks with and without admission control.}
  \label{fig:defense}
\end{figure}

\subsection{Admission Control to Defend Against PUE Attacks}

In CR network, due to the nature of dynamic spectrum access, ongoing SU
services may be forced to be drop in the presence of PUs.
When a CR network suffers from PUE attacks, the phenomenon of dropping will
be severely magnified, leading to the discontinuity of SU services.
A Guard Channel (GC) is a simple but effective approach to protect the
ongoing services in a wireless networks. In this paper, we propose a GC-based
admission control strategy to defend against PUE attacks.
Upon the arrival of PUs or PUE attackers (if not detected), the ongoing
services have to perform spectrum handoff. The handoff services need to
aquire new available channels to resume the transmissions.
Similar to conventional GCs, the proposed approach reserves a certain portion
of the available channels for the handoff services.
Once an SU needs a new channel for transmissions, it has to send a request
message to the cognitive BS, applying for an available channel. The cognitive
BS observes the remaining available channels. If the number of available
channels is larger than the reservation number, the SU is allocated a new
available channel. Otherwise, the SU's request is denied. The proposed GC
strategy takes into account of the existence of PUE attacks, and has
considered channel reservation for the PUE attacks. Hence, the dropping rate
caused by PUE attacks could be significantly alleviated.

Fig.~\ref{fig:defense} compares the feature-based PUEA detection \cite{Pu11}
without admission control and the feature-based PUEA detection with admission
control. The results show that the introduction of admission control is able
to significantly reduce the dropping rate. It is clear that the admission
control based mechanism can significantly reduce the dropping rate. For
instance, when $P_d=0.9$ and $\lambda_{\mathrm{EU}}=0.8$, the dropping rate
is about $1.8\times 10^{-3}$ without admission control while it is only about
$6.2\times 10^{-5}$ with admission control. Consequently, the proposed
admission control based scheme can applied as an efficient defense approach.

\section{Conclusion}

This paper focuses on the PUE attack security problem in CR networks. A comprehensive introduction to PUE attacks is presented and several technical challenges are discussed, including classification of attackers, conditions for successful PUE attacks, and impacts of PUE attacks on CR networks. After
that, a database-assisted detection approach is proposed to efficiently
discover PUE attacks. Multi-threshold fast energy detection and fingerprint-based location verification are integrated and driven by a two-level database.
In addition, an admission control based defense
approach is proposed to alleviate the impact of  PUE attacks on the
performance of CR networks. By reserving a portion of channels for the handoff services, the dropping rate induced by successful PUE attacks could be evidently reduced. Illustrative results demonstrate that the reported
detection and defense approaches are effective in discovering and defending
PUE attacks in CR networks.

\section*{Acknowledagement}

This research is partially supported by program of NSFC (grant
no.~U1035001), the Natural Science Foundation of
Guangdong Province (grant no.~8351009001000002),
the project 217006 funded by the Research Council of Norway, the European Commission FP7 Project EVANS (grant no. 2010-269323), and the European Commission COST Action IC0902, IC0905 and IC1004.

\begin{IEEEbiographynophoto}{Rong Yu}
[S¡¯05, M¡¯08] (yurong@ieee.org) received his Ph.D. from Tsinghua University, China, in 2007. After that, he worked in the School of Electronic and Information Engineering of South China University of Technology (SCUT). In 2010, he joined the Institute of Intelligent Information Processing at Guangdong University of Technology (GDUT), where he is now an associate professor. His research interest mainly focuses on wireless communications and networking, including cognitive radio, wireless sensor networks, and home networking. He is the co-inventor of over 10 patents and author or co-author of over 50 international journal and conference papers.
Dr. Yu is currently serving as the deputy secretary general of the Internet of Things (IoT) Industry Alliance, Guangdong, China, and the deputy head of the IoT Engineering Center, Guangdong, China. He is the member of home networking standard committee in China, where he leads the standardization work of three standards.
\end{IEEEbiographynophoto}

\begin{IEEEbiographynophoto}{Yan Zhang}
(yanzhang@simula.no) received a Ph.D. degree from Nanyang Technological
University, Singapore. Since August 2006 he has been working with
Simula Research Laboratory, Norway. He is currently a senior research
scientist at Simula Research Laboratory. He is an associate professor
(part-time) at the University of Oslo, Norway. He is a regional editor,
associate editor, on the editorial board, or guest editor of a number
of international journals. He is currently serving as Book Series
Editor for the book series on Wireless Networks and Mobile
Communications (Auerbach Publications, CRC Press, Taylor \& Francis
Group). He has served or is serving as organizing committee chair for
many international conferences, including AINA 2011, WICON 2010, IWCMC
2010/2009, BODYNETS 2010, BROADNETS 2009, ACM MobiHoc 2008, IEEE ISM
2007, and CHINACOM 2009/2008. His research interests include resource,
mobility, spectrum, energy, and data management in wireless
communications and networking.
\end{IEEEbiographynophoto}

\begin{IEEEbiographynophoto}{Yi Liu}
[S'09] (liuii5115@yahoo.com.cn) received his M.S. degree in physical
electronics from Hunan Normal University, China in 2007 and the Ph.D.
degree in signal and information processing from South China University
of Technology (SCUT), China in 2011.
His research interests include cognitive radio networks, cooperative
communications and intelligent signal processing.
\end{IEEEbiographynophoto}

\begin{IEEEbiographynophoto}{Stein Gjessing}
(steing@ifi.uio.no) is a professor of computer science in Department
of Informatics, University of Oslo and an adjunct researcher at Simula
Research Laboratory. He received his Dr. Philos. degree from the University of Oslo in 1985. Gjessing acted as head of the Department of Informatics for  4 years from 1987. From February 1996 to October 2001 he was the chairman
of the national research program ``Distributed IT-System'', founded by the
Research Council of Norway. Gjessing participated in three European funded
projects: Macrame, Arches and Ascissa. His current research interests are routing,
transport protocols and wireless networks, including cognitive radio and
smart grid applications.
\end{IEEEbiographynophoto}

\begin{IEEEbiographynophoto}{Mohsen Guizani}
[FIEEE'09] is currently a Professor and Associate Vice President for
Graduate Programs at Qatar University. He served as the Vice Dean of
Academic Affairs at Kuwait University. Previously, he was the Chair of
the CS Department at Western Michigan University from 2002 to 2006 and
Chair of the CS Department at the University of West Florida from 1999
to 2002. He also served in academic positions at the University of
Missouri-Kansas City, University of Colorado-Boulder, and Syracuse
University. He received his B.S. (with distinction) and M.S. degrees in
Electrical Engineering; M.S. and Ph.D. degrees in Computer Engineering
in 1984, 1986, 1987, and 1990, respectively, from Syracuse University,
Syracuse, New York. His research interests include Computer Networks,
Wireless Communications and Mobile Computing, and Optical Networking.
He currently serves on the editorial boards of six technical Journals
and the Founder and EIC of "Wireless Communications and Mobile
Computing" Journal published by John Wiley
(http://www.interscience.wiley.com/jpages/1530-8669/). He is also the
Founder and the Steering Committee Chair of the Annual International
Conference of Wireless Communications and Mobile Computing (IWCMC). He
is the author of seven books and more than 270 publications in refereed
journals and conferences. He guest edited a number of special issues in
IEEE Journals and Magazines. He also served as member, Chair, and
General Chair of a number of conferences.
Dr. Guizani served as the Chair of IEEE ComSoc WTC and Chair of TAOS
ComSoc Technical Committees. He was an IEEE Computer Society
Distinguished Lecturer from 2003 to 2005.
Dr. Guizani is an IEEE Fellow and a Senior member of ACM.
\end{IEEEbiographynophoto}

\end{document}